\documentclass{llncs}
\usepackage{llncsdoc}
\usepackage{graphicx} 
\begin{document}
\thispagestyle{empty}
\pagestyle{empty}
\title{Cellular automata with majority rule on evolving network}

\author{Danuta Makowiec\inst{} }

\institute{Institute of Theoretical Physics and Astrophysics,
Gda\'nsk University,\\  80-952 Gda\'nsk, ul. Wita Stwosza 57, Poland}

\maketitle

\begin{abstract}
The cellular automata discrete dynamical system is considered as the two-stage process: the majority rule for the change in the automata state and the rule for the change in topological relations between automata. The influence of changing topology to the cooperative phenomena, namely zero-temperature ferromagnetic phase transition, is observed.\end{abstract}

\section{Introduction}
Lattice models are the basic instrument in the study of phase transitions in equilibrium statistical mechanics (see, e.g., \cite{StatMech}). The Ising model is the simplest model of nearest-neighbor ferromagnetic interactions where the collective features are studied. Traditionally, the phase transition has been considered in regular lattices. According to the  renormalization group theory, detailed structure of interactions as well as the structure of network connections are irrelevant. Therefore, for example, cellular automata have been continuously tested with the hope that these systems can imitate nature \cite{CA_ferro}.

If the cellular automata are considered on a regular lattice and  each site $i$ of the lattice is occupied by a binary variable, $s_i$, which assumes the values $\pm 1$, then referring to these variables as {\it spins} which point either `up' ($+1$) or `down' ($-1$), we employ the terminology of the Ising model. The Ising nearest-neighbor interaction in case of cellular automata becomes usually a simple majority rule, it is: if most of the nearest neighbors of $i$ spin point up (down) at time $t$ then the value of the $i$ spin at time $t+1$ is up (down). This deterministic rule is interpreted as the zero-temperature approximation to the Ising interactions. The research here is aimed on the ergodicity, i.e., on the uniqueness of the final state. Especially,  one asks if starting from a random distribution of spins states one ends up with a homogeneous states of {\it all spins up} or {\it all spins  down}. To mimic the temperature effects one adds a stochastic noise to perturb with some probability the execution of the deterministic dynamics.  The research here is focused on properties of the transition from the non-ordered  phase to the ordered phase - ferromagnetic phase transition.

Recent research in the structure and topology of real networks \cite{BA,NewmanReview} has shown that social, biological, technological networks are far from being regular.  However, they also are  being  far from a random network.  It has been  shown that networks called {\it small-world} networks and  {\it scale-free} networks exhibit mixed properties of regular lattices and a random graph \cite{BA,WS}. This has trigged to study of standard models of statistical mechanics in these networks. The transition from non-ordered phase to the ordered phase in the spin system spanned on the small-world network shows a change in behavior from the regular case to the mean-field characteristics \cite{Gitterman,Svenson}. Interesting that, depending on the network construction, the phase transition at the finite temperature exists or not \cite{NovotnyWheeler}. Moreover, when the  scale-free network is employed then the critical properties of the Ising model are different  from those ones observed in the regular networks \cite{Aleksiejuk,Goltsev}. Many authors have considered other problems in these new networks: percolation properties \cite {MooreNewman}, the spread of infection diseases \cite{infection} social relations \cite{social}, computer organization \cite{computer}.

Small-world networks are intermediates between regular lattices and the random graph. A small-world network is generated by rewiring with a probability $p$ the links of a regular lattice by long-distance random links \cite{WS}. The presence of a small fraction of `short-cuts' connecting otherwise distant points, drastically reduces the average shortest distance between any pair of nodes in network.

In case of cellular automata such the small-world network means that the cellular automata is no longer homogeneous --- inhomogeneity is due to the topology. Sets of nearest-neighbors differ from a site to a site.

There is one important property which is ubiquitous in biological and artificial networks and which is missed in the small-world network ---  the  distribution of the vertex degree does not show wings decaying as a power-law. Two important ingredients have been shown to be sufficient to generate such feature: growing number of vertices and preferential attachment of links. The well established Barabasi-Albert model based on these two mechanisms  has been proposed \cite{BA}. The  distributions of the vertex degree in this network is of the form $P(k) \propto k^{-\gamma} $ where $\gamma \in [2,3]$ and $k$ a vertex degree. Thus adapting the Barabasi-Albert idea to cellular automata one obtains a computing system with the topology that evolves.

In this paper we address the question of the role played by the topology in the zero-temperature ferromagnetic transition considered in the  cellular automata of spins. The starting system is the cellular automata on a square lattice.  The square lattice used here is strongly clustered in the sense that if $j$ and $k$ are neighbors of $i$ then there is a short path between them that does not pass through $i$. In order to keep the high clustered property (typical for the small-world network) and reproduce the distribution of the vertex degree with the power-low decay (sign of the scale-free network) we propose the evolution rule for the network. Hence, the cellular automata time step means the change of the spin states and the change of the network topology. It will be shown that with increasing $p$- the rewiring parameter, the cellular automata work as the solver to the {\it density classification task} \cite{Mitchell}, highly accurately, namely our cellular automata converge to a fixed point of all 1's if the initial configuration contains more 1's than -1's by 2\%, and, symmetrically,  to a fixed point of all -1's if the initial configuration contains more -1's than 1's by 2\%.

\section{The model description}

We present the model of the  spin cellular automata located in $N$ vertices among which there is an explicit integer parameter $N_0$ fixing the total number of edges.   These $N$ spins are located to form  a square lattice with the periodic boundary conditions. Hence, if the linear size of the lattice is $L$ then $N=L\times L$ and $ N_0=4N$. We start with the ordinary nearest-neighbor relations, i.e., for any $i$ site  the nearest-neighbor spins are located in the following  set of vertex indices: $N(i,0)=\{ i-L, i-1, i, i+1, i+L \} $. Initially, each spin state is set up  randomly with the probability $\rho$.

The evolution step  consists of the two subsequent steps: (A) the asynchronous stochastic evolution of topological relations and (B) the synchronous deterministic majority rule.

\subsection{Evolution of the lattice}

In each time-step $t$ and for each vertex  $i$ the following rearrangement of the edges is performed: from the set of neighbors $N(i,t)$ of the vertex $i$ (at the given time $t$) one element $j$ is chosen at random with the probability $p$. Thus $p$ relates to  the parameter of the small-world evolution \cite{WS}. Then, another vertex  $k$ is picked up from the set of all vertices $N-\{ i\}$ randomly. A new edge is created between $i$ and $k$ independently of the fact that there was or not an edge between them. The edge between $i$ and $j$ is deleted. Hence the total number of edges $N_0$ is conserved. This  evolution we will call stochastic on the contrary to the evolution in which some preferences are applied. The proposed preferences are in the agreement with the basic conviction how to obtain a scale free distribution for the vertex degree: `the richer you are, the richer you get` --- principle \cite{BA}.

The preferences are as follows:  in each time step $t$ and for each vertex $i$, from the set of neighbors $N(i,t)$ an element  $j$ is chosen with the probability modified by the degree of $j$ vertex. Namely, the probability that $j$ neighbor is unlinked is
\begin{equation}
p[1-{deg(j)\over T}]
\label{eq1}
\end{equation} 
where $deg(j)$ denotes the degree of $j$ vertex, $T$ denotes the value of the unlink threshold. The degree of a vertex denotes the number of edges attached to the vertex. By (\ref{eq1}) if $deg(j) \ge T$ then it is impossible to unlink $j$. This process will be called the intensional detachment. The edge between $i$ and $j$ is deleted. Then a new vertex is randomly chosen but with the preference to link to a vertex with the high vertex degree. It is, a randomly picked up vertex $k$ is accepted with probability 
\begin{equation}
{deg(k)\over T}
\label{eq2}
\end{equation} 
Thus if $deg(k) \ge T$ then randomly chosen $k$ is certainly accepted.
This process will be called the preferential attachment. 

In the following we will name:\\
{\it model 0} : the completely stochastic edge evolution,\\
{\it model 1} : the intensional detachment and the stochastic attachment,\\
{\it model 2} : the stochastic detachment and the preferential attachment,\\
{\it model 3} : the intensional detachment and the preferential attachment.

\subsection{Evolution of spins}
Synchronously, in each time step,  the majority rule is applied to set the state of every spin in the next time-step. The rule acts on all spins belonging to the the set of nearest neighbors $N(i,t)$ of the given $i$th spin. In the case when the result of voting is zero, the $i$th spin does not change the state.

\section{Results}
The system simulated is a square lattice with $L=200$ and periodic boundary conditions. We start with a random distribution of spin states with probability $\rho$ to set the spin state to  $+1$. In each time step we measure {\it magnetization} -- the sum of all spins states normalized by $L^2$. If the magnetization is $1$ or the time step reaches the limit of the  hundredth step, then we stop the evolution and record the following characteristics: magnetization, susceptibility (the variation of the magnetization),  vertex degree and  number of time steps to reach the stable state. The experiment is repeated 200 times for each $p$ and $\rho$ value. The threshold value for the preferences, see (\ref{eq1}), (\ref{eq2}),  is $T=8$ in all experiments.

\subsection{Final state characteristic}

\begin{figure}
\includegraphics[width=0.50\textwidth]{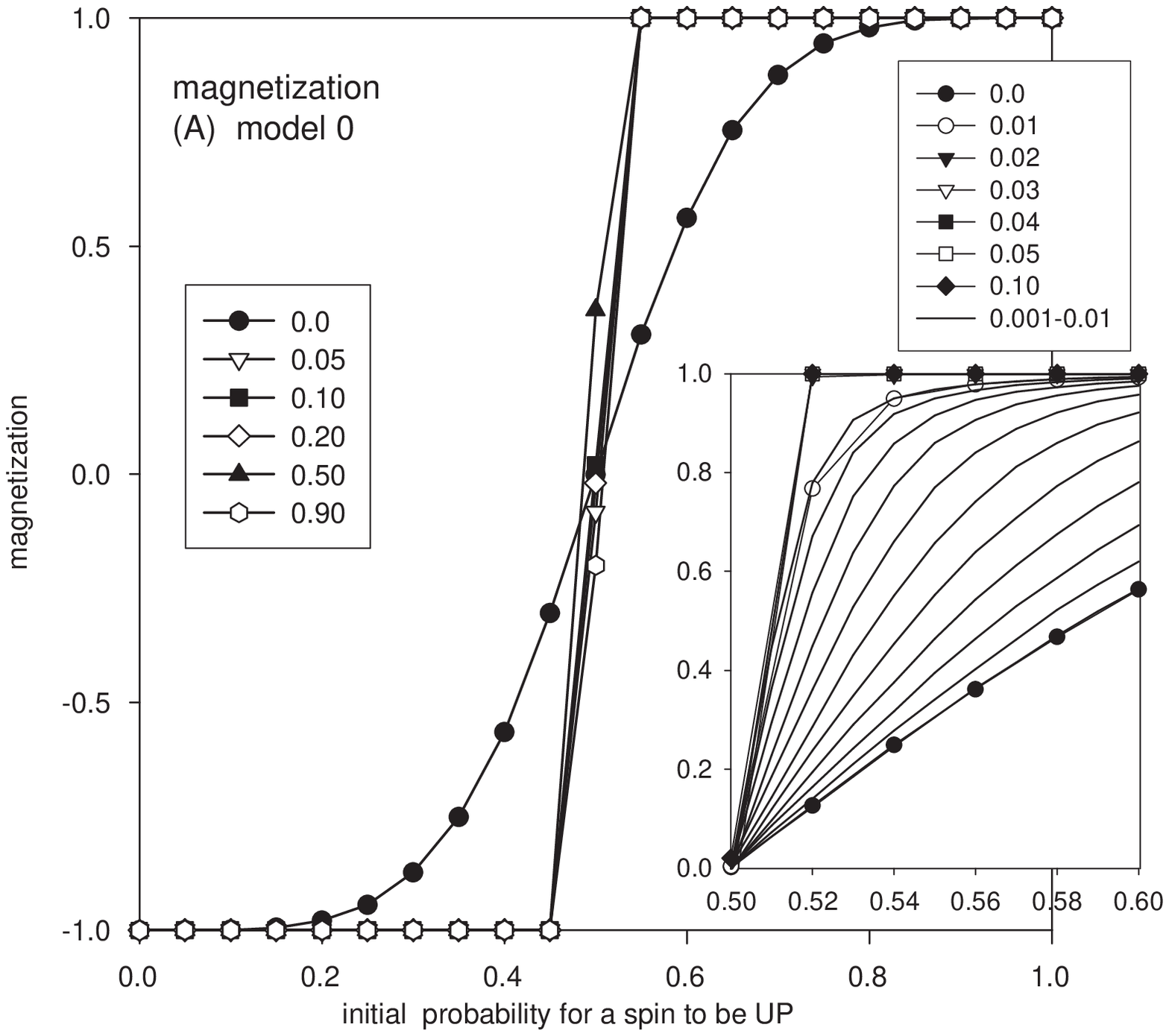}
\includegraphics[width=0.50\textwidth]{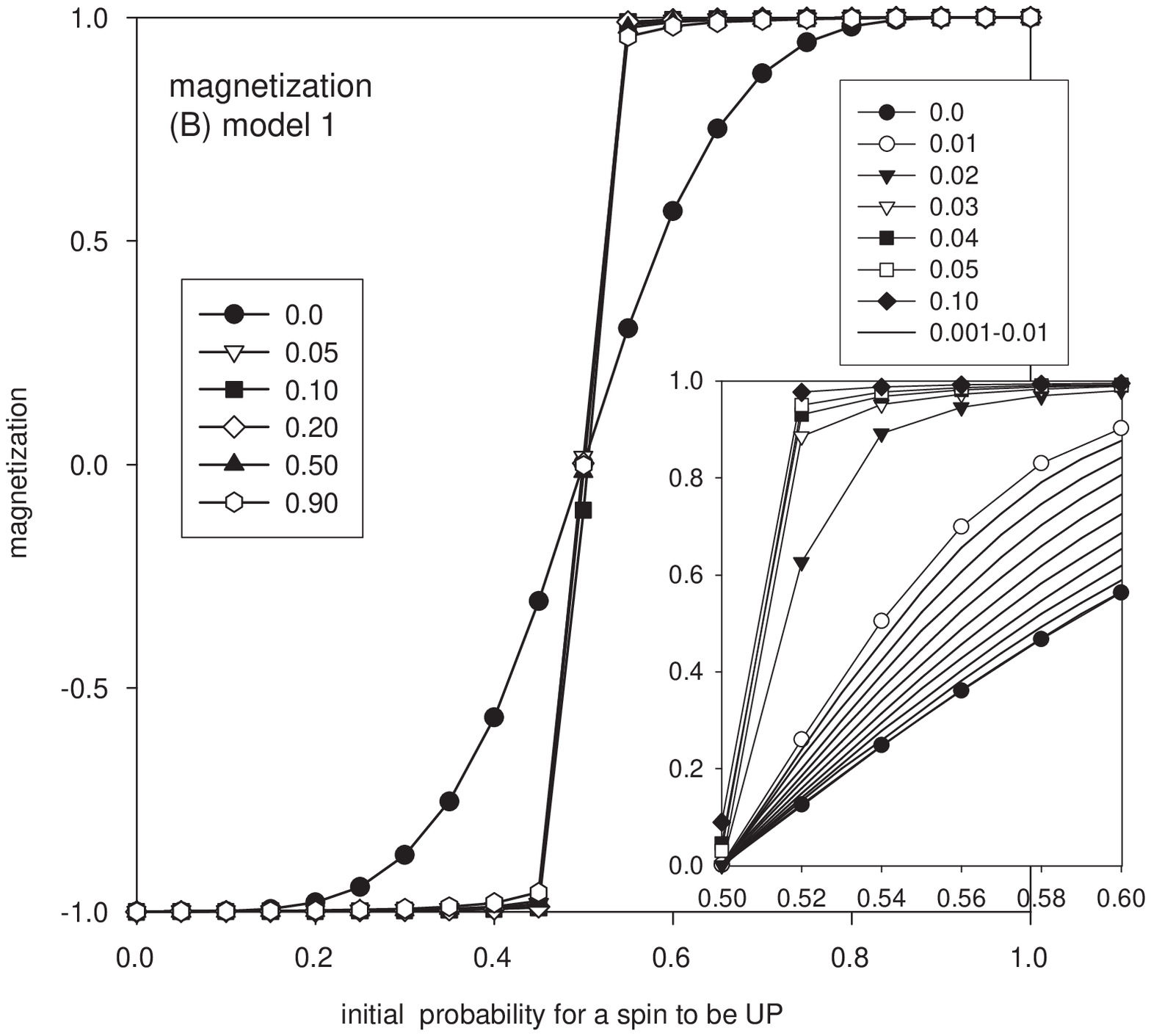}\\

\includegraphics[width=0.50\textwidth]{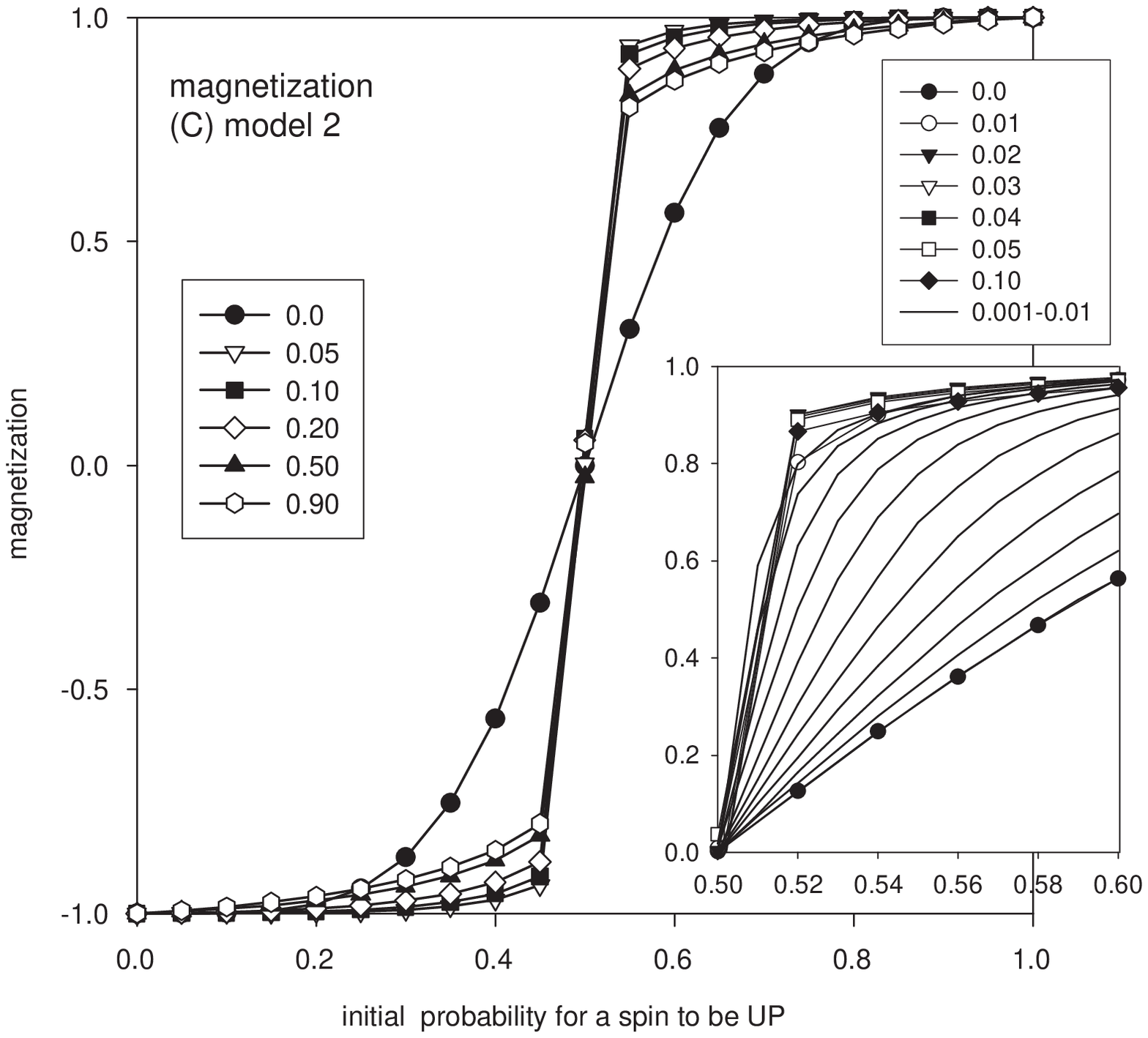}
\includegraphics[width=0.50\textwidth]{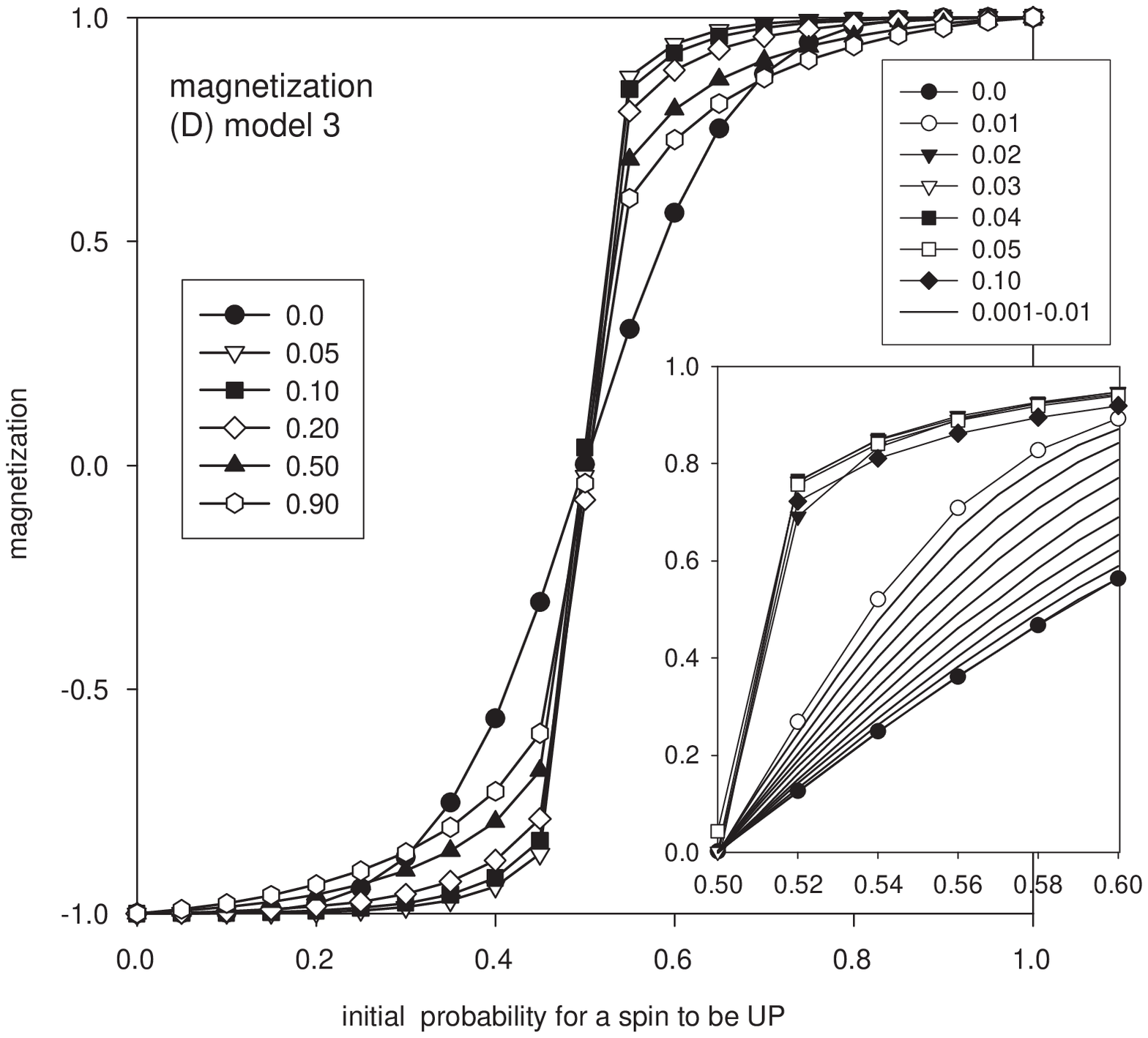}
\caption[]{\label{magn}Magnetization of final configurations vs initial probability for a spin to be UP ($\rho$) for different values of short-cut edges ($p$) what is indicated by curve's labels: (A) model 0 -- stochastic linking and unlinking, (B) model 1 -- intentional unlinking, (C) model 2 -- preferential linking to, (D) model 3 -- intentional unlinking and preferential linking to. The large frames present the general dependences on $p$ and $\rho$; the inside frames show the change of magnetization if $\rho$ is close to 0.5 and $p$ is small. The lines without point markers are results for $p= 0.001, 0.002,\dots, 0.01$ subsequently.}
\end{figure}

\begin{figure}
\includegraphics[width=0.50\textwidth]{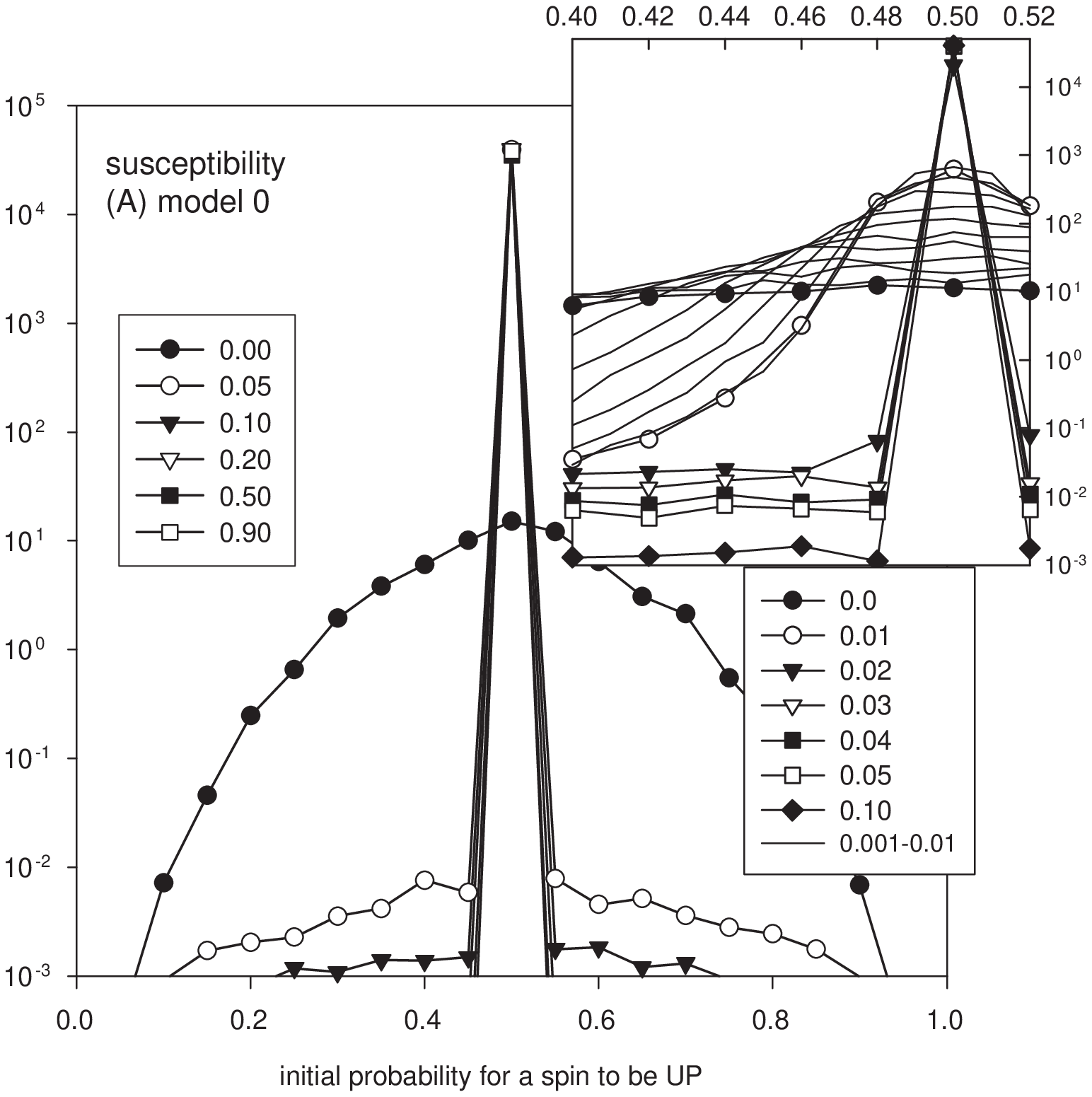}%
\includegraphics[width=0.50\textwidth]{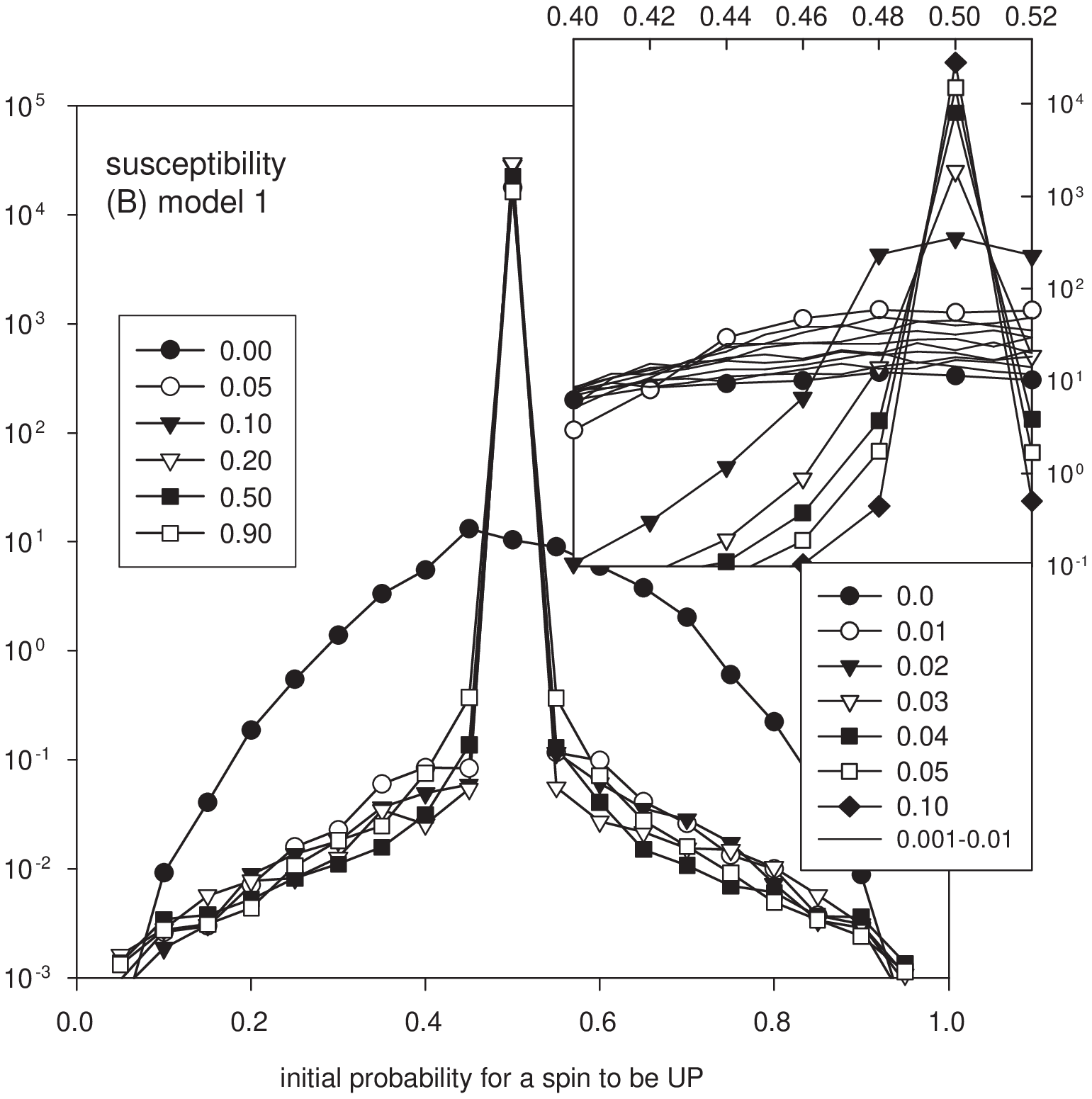}%

\includegraphics[width=0.50\textwidth]{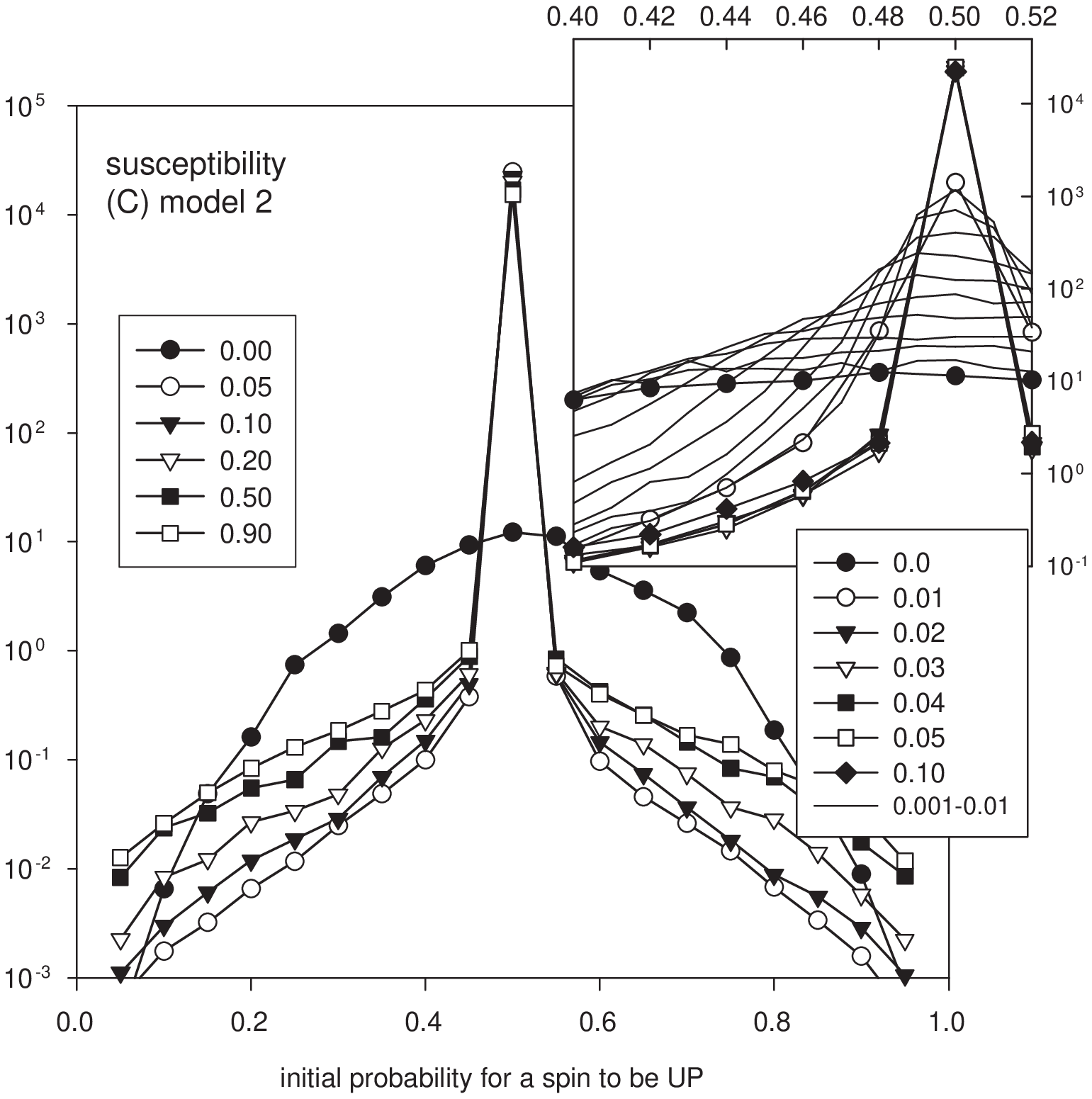}%
\includegraphics[width=0.50\textwidth]{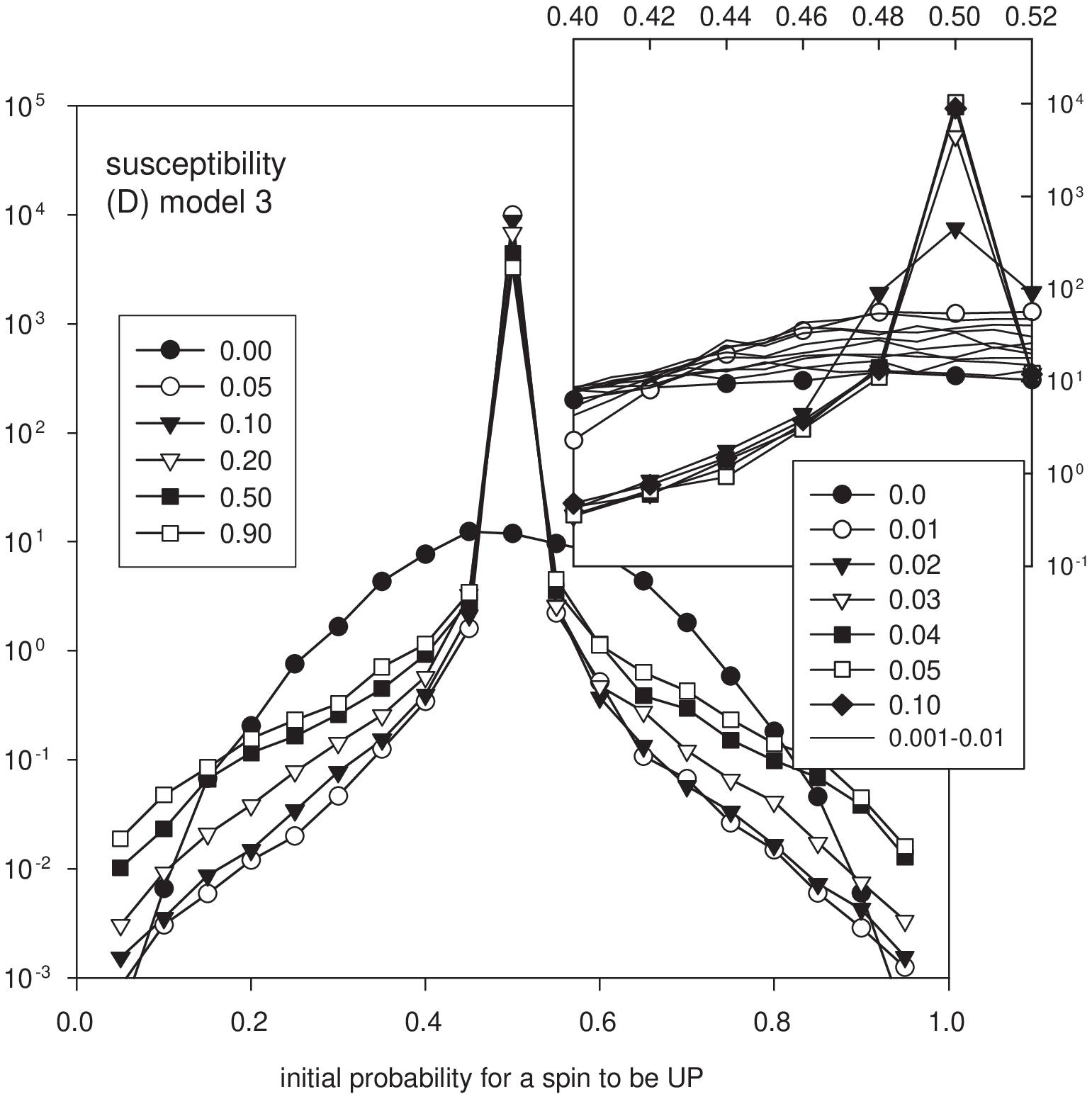}%
\caption[]{\label{susc}Log plots of susceptibility of final configurations vs initial probability for a spin to be UP ($\rho$) for different values of short-cut edges ($p$) what is indicated by curve's labels: (A) model 0 -- stochastic linking and unlinking, (B) model 1 -- intentional unlinking, (C) model 2 -- preferential linking to, (D) model 3 -- intentional unlinking and preferential linking to. The large frames present the general dependences on $p$ and $\rho$; the inside frames show the change of susceptibility if $\rho$ is close to 0.5 and $p$ is small. The lines without point markers are results for $p= 0.001, 0.002,\dots, 0.01$ subsequently.}
\end{figure}

The sharp change in properties of the final state of cellular automata  is observed when $\rho$ is crossing $0.5$, see Fig.\ref{magn}, Fig.\ref{susc}. The final state switches between the state of {\it all spins down} if $\rho <0.5$ (magnetization $=-1$) to the state of {\it all spins up} if $\rho >0.5$ (magnetization $=1$), Fig.\ref{magn}. Depending on the model of the network evolution this transformation goes fast: model 0, or less rapidly: model 3. 
The transformation can be estimated by measuring the intervals of $\rho$ around the 0.5- value which lead to the final states with with magnetization close to $\pm 1$, see Fig.1. More accurate marks for these  intervals can be found by observing susceptibility, see Fig.2. In case of unperturbed network ($p=0$) the susceptibility takes the parabola-like shape on the log-plots. With increasing $p$ the suscpetibility becomes the sharply pointed curve with maximum at $\rho=0.5$. The basic step in $\rho$ in the experiments performed is $\Delta \rho=0.02$. Therefore the peaks of susceptibilities are recognized within the interval $(0.48, 0.52)$. In the small windows of Fig.2 one can observe how fast and which way the parabola transforms into the sharply pointed curves. The significant qualitative change occurs when $p$ changes from $p=0.01$ to $p=0.02$. In particular, if one assumes that the interval of critical changes means the susceptibility  greater than 1 (at average this condition is equivalent to the demand of the absolute magnetization being smaller than $0.9$), then one finds:\\
--- $\rho \in (0.30, 0.70) $ in case of cellular automata without the network evolution\\
--- $\rho \in (0.48, 0.52) $ in case of cellular automata with the network evolving according to the rules of models 0,1,2\\
--- $\rho \in (0.46, 0.54) $ in case of cellular automata with the network evolving according to the  rule of model 3.\\ 
Because of the observed rapid changes in main characteristics one can say that the cellular automata solve  the density classification task. Especially, model 0 solves the density classification problem  extremely efficiently: quickly and with the high certainty.
  
Also time needed to reach the fixed point stabilization changes when $p$ is increasing. If $p>0.1$ and $\rho << 0.5$ then the time to stabilization is larger than in case of the system with not evolving network. However, the stabilization is reached in less than 40 steps.  If $\rho \in (0.4,0.6)$  and $p$ is small,  $p<0.01$, then the fixed point stabilization is not observed in less than 100 steps. The properties shown in figures collect features found as snap-shots of cellular automata states at the one hundredth step. This limitation does not introduce any important restriction. It is because that when the network does not evolve then, due to the many invariants of the spin-state rule,  one observes the stabilizations on configurations which oscillate with some time period. This results in that the stabilization time is read as $t=100$. Such limit configurations occur also if $p$ is small, i.e., $p<0.01$. When the rewiring process is stronger the oscillating patterns disappear in less than 100 steps. The stabilization different from the fixed points is not observed if $p > 0.01$.

\subsection{Degree distribution} 
\begin{figure}
\includegraphics[width=0.50\textwidth]{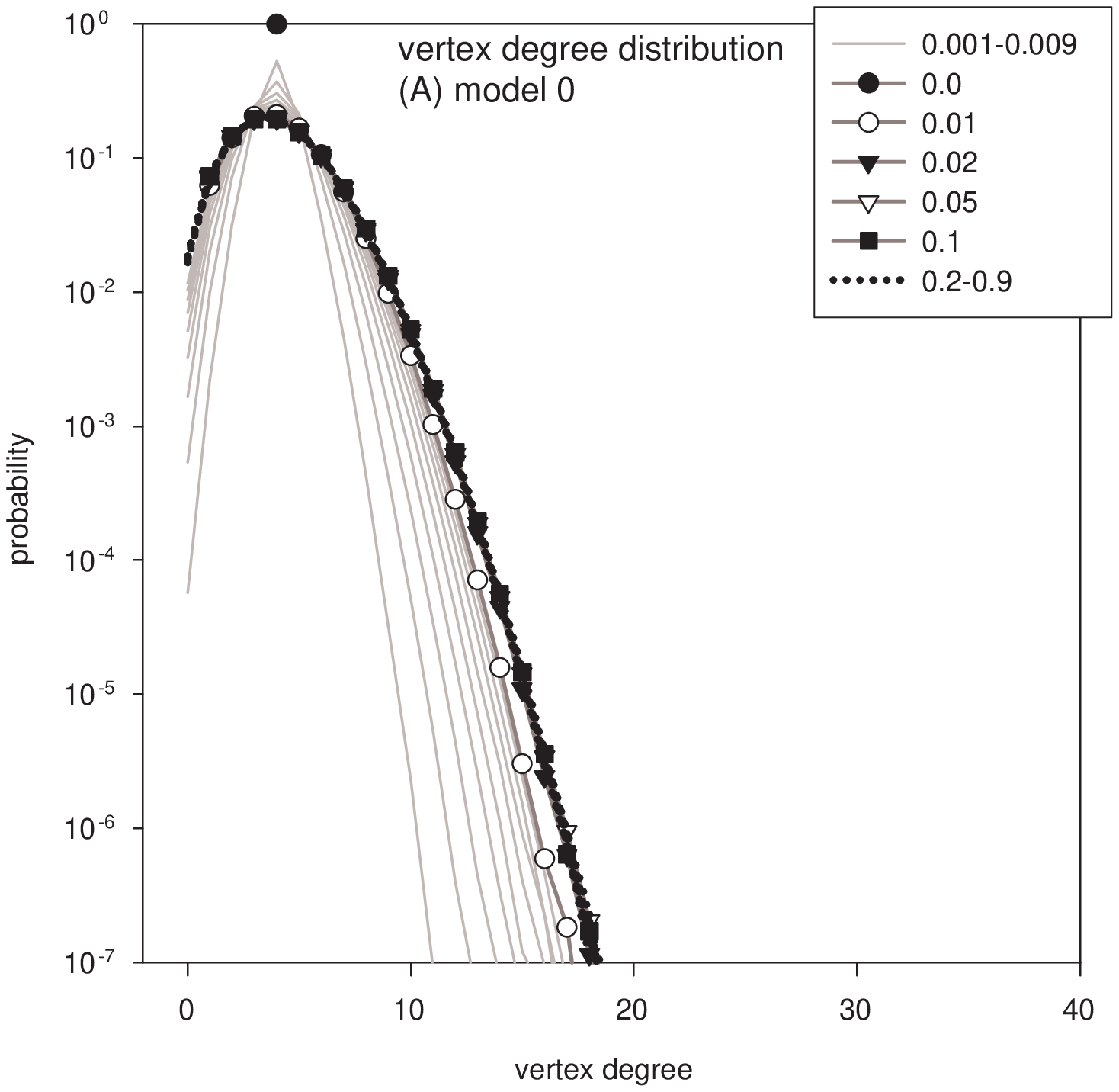} \includegraphics[width=0.50\textwidth]{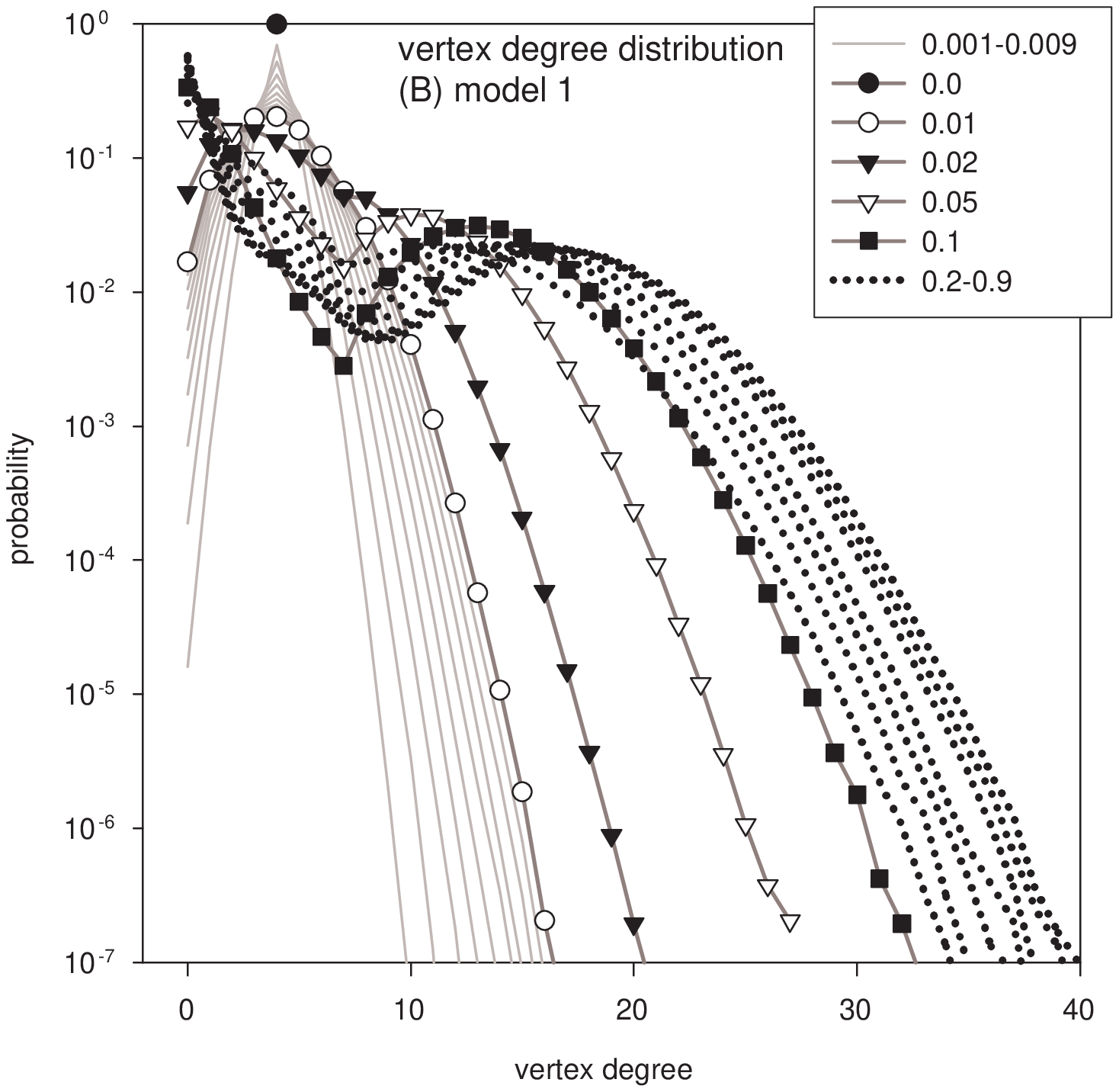}

\includegraphics[width=0.50\textwidth]{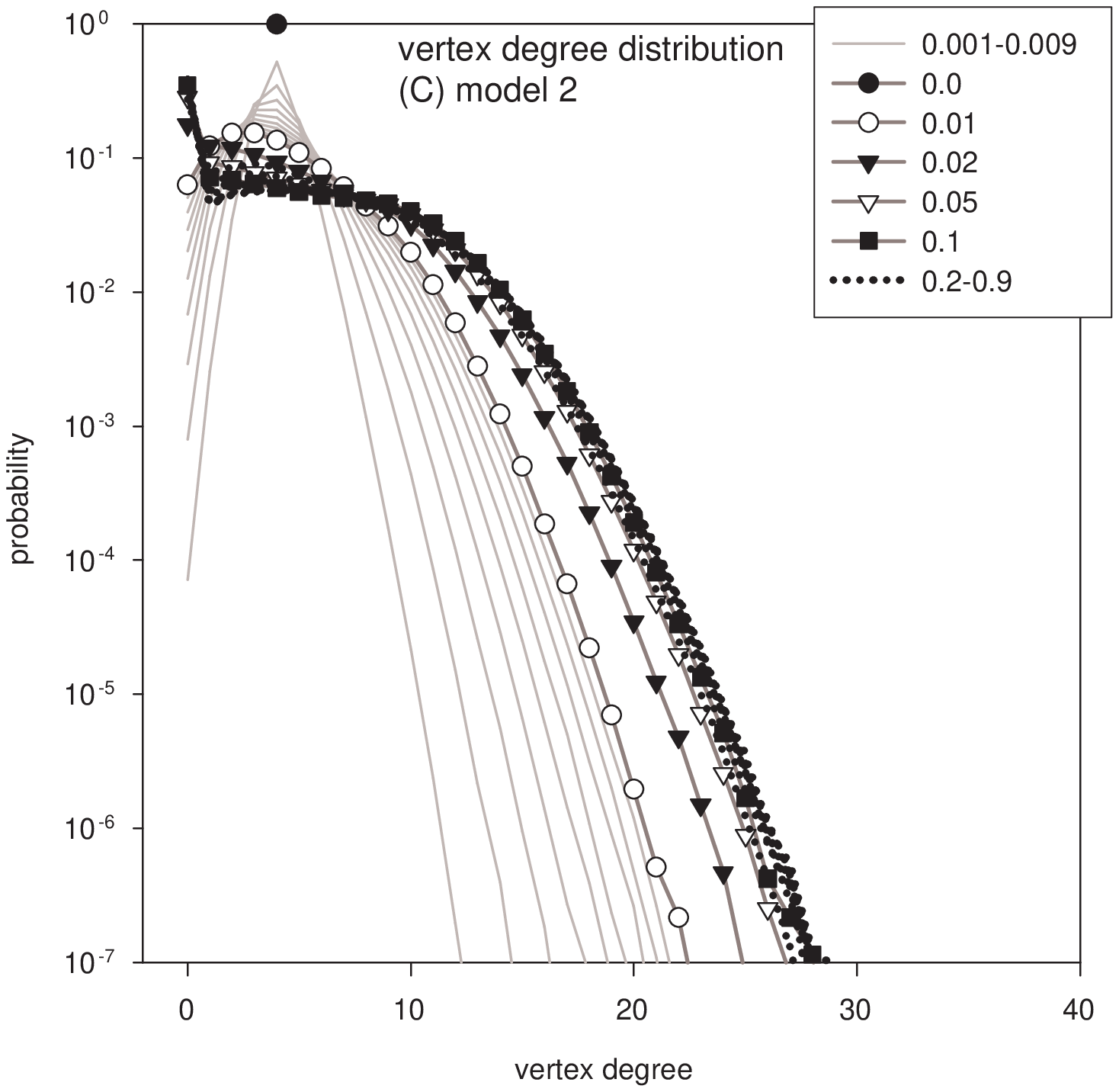}%
\includegraphics[width=0.50\textwidth]{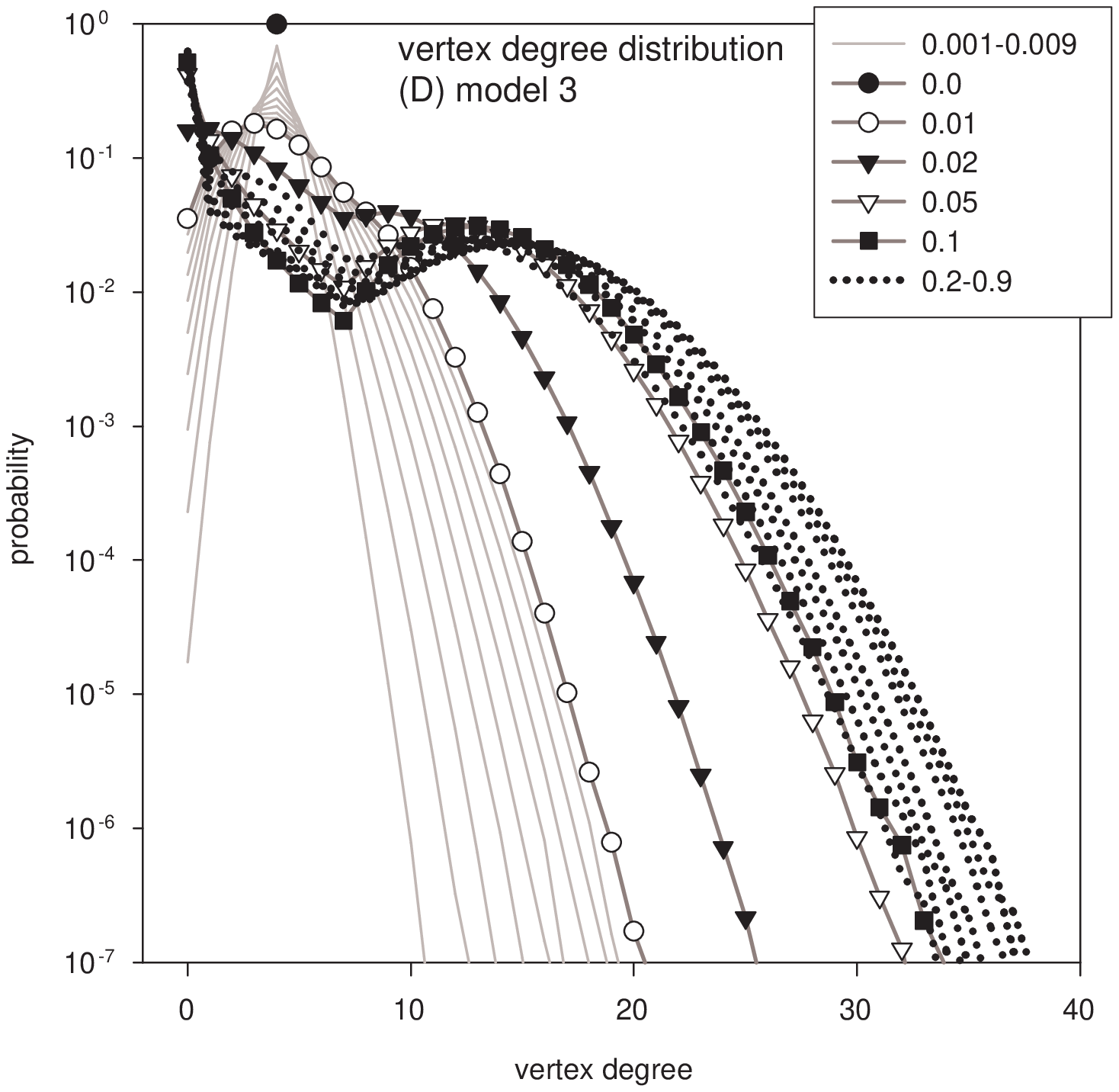}%
\caption[]{\label{degree}Log plots of vertex degree in final configurations vs initial probability for a spin to be UP ($\rho$) for different values of short-cut edges ($p$) what is indicated by curve's labels: (A) model 0 -- stochastic linking and unlinking, (B) model 1 -- intentional unlinking, (C) model 2 -- preferential linking to, (D) model 3 -- intentional unlinking and preferential linking to. The gray lines  are results for $p= 0.001, 0.002,\dots, 0.009$ subsequently; the dotted lines are results for $p= 0.2, 0.3,\dots, 0.9$ subsequently.}
\end{figure}

The vertex  degree distributions are presented in Fig.\ref{degree}. The plots collect results recorded when the cellular automata systems arrive to the limit of either a ferro fixed point stabilization or a hundred step of time. Hence the total number of changes in the network depends on the time given to the evolution. However, if $p\ge 0.01 $ then after a hundredth steps the average probability for an edge to change is 1. Thus, the difference between a network state after a hundred steps with $p=0.01$ and after the first step with $p=1$ consists in the synchronousness of events: with increasing $p$ the edge evolution becomes more synchronous.

All plots of the vertex degree distribution with $p \le 0.01$ characterize the network configurations that are obtained at the one hundredth step. In case of model 0 the fast convergence to the Gaussian distribution (the parabola shape in the log-plot) centered at $k=4$ is observed. The significant distinction from the Gaussian shape in cases of other models is due to the preferences introduced. One observes  vertices with the degree higher than 20 which in the Gaussian world do not exist. Because of the threshold value considered here, $T=8$, the vertices with the degree larger than 8 are protected from unlinking and preferred in linking. Therefore  the number of vertices with high degree is growing with increasing $p$. However, the presented  distributions do not have wings depending polynomially  on the vertex degree, namely,  decaying as $k^{-\gamma}$ for some $\gamma$ and $k$ vertex degree .

\section{Concluding remarks}

We have considered properties of spin cellular automata initially spanned on the regular square lattice and then rewired each time step systematically. What we have observed is the final spin state. We have concentrated on the two configurations: {\it all spins up} and { \it all spins down}. It has appeared that only these two configurations are possible as the final configuration for spins. Which one of these two emerges is depended on $\rho$ the density of $1$ in the initial configuration. Due to the changes in the network the interval of uncertainty about the final configuration contracts sharply to the small interval around the critical value of $\rho=0.5$. From the statistical physics point of view  main investigations concern the ergodicity problem, i.e. the memory about the initial state properties in the evolving systems. The strong connectivity arisen from the small-world network enhances sharply the solution. 
The cellular automata considered are ergodic outside the interval $\rho\in(0.48,0.52)$  what means that the interval of ergodicity is  definitely larger  than the interval of $\rho$ in case of the ordinary spin cellular automata. Therefore we claim that the systems considered can work as solvers of the density task.
Moreover, since the analysed properties have been collected at time moments much smaller than the system size, the stated features seem to be universal in the sense that they are size independent.

Our work is only preliminary. The dynamical system considered here needs further investigations. We have studied  the zero-temperature Ising problem. Now the study of stochastic spin evolution should be undertaken. Moreover, the networks considered here can bee included into the one thermodynamic ensemble, so-called canonical ensemble, because all networks have the same number of edges \cite{Farkas}. For the canonical ensemble there is a possibility to fix a temperature and associate the energy to every network configuration. Such approach states the new question: how this temperature and this energy relate to the ordinary thermodynamic characteristics of spin system.

{\bf Acknowledgments}\\
We wish to acknowledge the financial support of Polish Ministry of Scientific Research and Information Technology : PB$\slash$1472$\slash$PO3$\slash$2003$\slash$25

\end{document}